\documentclass[prl,aps,superscriptaddress,floatfix,tightenlines,showpacs,twocolumn]{revtex4}

\usepackage{graphicx}
\usepackage{dcolumn}   
\usepackage{bm}        
\usepackage{amssymb}   
\usepackage{amsmath}
\usepackage{url,CJK}
\usepackage{layout,color}
\usepackage{epsfig}
\usepackage{graphicx}
\usepackage{epstopdf}
\usepackage{booktabs}
\usepackage{float,braket}
\usepackage[hyperindex,breaklinks]{hyperref}

\def\be{\begin{equation}}
\def\ee{\end{equation}}
\def\ba{\begin{eqnarray}}
\def\ea{\end{eqnarray}}

\begin{document}
\begin{CJK}{UTF8}{gbsn}
\title{Construction of Maximally Localized Wannier Functions}
\author{Junbo Zhu}
\affiliation{International Center for Quantum Materials, School of Physics, Peking University, Beijing 100871, China}
\author{Zhu Chen}
\affiliation{Institute of Applied Physics and Computational Mathematics, Beijing 100088, China}
\author{Biao Wu}   \email{wubiao@pku.edu.cn}
      \affiliation{International Center for Quantum Materials, School of Physics, Peking University, Beijing 100871, China}
      \affiliation{Collaborative Innovation Center of Quantum Matter, Beijing 100871, China}
      \affiliation{Wilczek Quantum Center, College of Science, Zhejiang University of Technology,  Hangzhou 310014, China}
      \affiliation{Synergetic Innovation Center for Quantum Effects and Applications (SICQEA), Hunan Normal University, Changsha 410081, China}

\begin{abstract}
We present a general method of constructing maximally localized Wannier functions.
It consists of three steps: (1) picking a localized trial wave function, (2) performing 
a full band projection,  and (3) orthonormalizing with the L\"owdin method. 
Our method is capable of producing maximally localized Wannier functions 
without further minimization, and it  can be applied straightforwardly to random 
potentials without using supercells. The effectiveness of our method is 
demonstrated for both simple bands and composite bands.
\end{abstract}
\date{\today}
\maketitle
\end{CJK}

Wannier functions that are localized at lattice sites are an alternative representation
of electronic states in crystalline solid to the Bloch wave representation~\cite{Wannier37}.
They  offer an insightful picture of chemical bonds,  play a pivotal role  in the modern
theory of polarization~\cite{Resta94,Vanderbilt93}, and are the basis for an
efficient linear-scaling algorithms in electric-structure calculations~\cite{Goedecker99,Galli96}.
Wannier functions are also important in linking cold atom experiments in continuous
light potentials with lattice Hamiltonians, such as the Bose-Hubbard model and Anderson
random lattice~\cite{Jaksch98,White09,Ceperley10}.

Although there are explicit formula that transform  Bloch waves to  Wannier functions,
the construction of  Wannier functions is far from trivial. The primary reason is that
there are infinite number of Wannier functions for a given band due to  phase
choices for Bloch waves whereas in practice the maximally localized Wannier functions
(MLWFs) are sought and preferred~\cite{Marzari97}.
For one dimensional lattice, Kohn showed how to fix the Bloch wave phases to obtain these
MLWFs~\cite{Kohn59}.  In 1971, Teichler found a general method to construct
Wannier functions, which is insensitive to the phases of Bloch waves~\cite{Teichler}.
However, this method does not guarantee maximal localization and depends on initial trial
function.  A more sophisticated method involving numerical minimization 
of the spread of a Wannier function has been
developed to compute MLWFs~\cite{Marzari12}. There are also some
new developments recently~\cite{Nenciu2016,Louie2016,Levitt}.

In this work we present a general method for computing MLWFs.
Our method is somehow similar to Teichler's method as it also involves projection
and the L\"owdin orthonormalization method~\cite{Lowdin}. Nevertheless there is
a significant difference so that our method can produce MLWFs and is insensitive
to the initial trial wave functions. Compared to the
method in Ref. ~\cite{Marzari12}, our method does not need the minimization procedure.
As our method uses the full band projection, it can be applied 
straightforwardly to random potential without using supercells.
The effectiveness of our method is demonstrated for both simple bands and composite
bands.

We consider a simple band that is isolated from other bands.
Composite bands will be discussed later. Our method of constructing MLWFs
consists of three steps:
\begin{enumerate}
\item Guess: choose a set of trial wave functions $\ket{g_n}$, which are localized at lattice sites.
\item Projection: $\ket{\xi_n}=P\ket{g_n}$, where $P=\sum_k \ket{\psi_k}\bra{\psi_k}$ with
the summation over the whole first Brillouin zone.  Note that $P$ is a full band projection and
is different from the projection in Refs.~\cite{Marzari97, Teichler}, which is at a given
$k$ point. As a result of this crucial difference, the projected function $\ket{\xi_n}$ is
localized  at site $n$.
\item Orthonormalization: use the L\"owdin orthonormalization method~\cite{Lowdin}
to transform $\ket{\xi_n}$ into a set of MLWFs $\ket{w_n}$.  If one uses other methods 
such as Kohn's method~\cite{Kohn1973orth} to orthonormalize $\ket{\xi_n}$, 
the resulted Wannier function is unlikely maximally localized.
\end{enumerate}

Here is  why our method is effective and capable of producing
MLWFs. We re-write the full band projection operator in terms of Wannier functions,
$P=\sum_n\ket{w_n}\bra{w_n}$, where the summation is over all lattice sites.
We thus have
\begin{equation}
\ket{\xi_n}=\sum_m\ket{w_m}\braket{w_m | g_n}\approx
\sum_{m=\langle n\rangle}\ket{w_m}\braket{w_m | g_n}\,,
\end{equation}
where $\langle n\rangle$ indicates that the summation is only over site $n$ and
its nearest neighbors. It is clear that $\ket{\xi_n}$ is localized at lattice site $n$. However, 
these projected functions $\ket{\xi_n}$'s are not orthonormal.  Any  
orthonormalization of $\ket{\xi_n}$'s will give us a set of Wannier functions. 
For example, one may use Kohn's method~\cite{Kohn1973orth}.
We choose the L\"owdin method~\cite{Lowdin} as it can produce the MLWFs.  
According to Ref.~\cite{Aiken},
the L\"owdin orthogonalization uniquely minimizes the functional measuring the least squares
distance between the given orbitals and the orthogonalized orbitals. In our case,
for the set of projected orbitals $\ket{\xi_n}$,  the L\"owdin method produces 
Wannier functions $\ket{w_n}$ that minimize 
\begin{equation}
\label{sum}
\sum_n\int dx |\braket{x|\xi_n}-\braket{x|w_n}|^2\,,
\end{equation}
where the summation is over all lattice sites. For a periodic potential, this is 
equivalent to maximizing
$\braket{w_n | \xi_n}=\braket{w_n | g_n}$. Therefore, if $\ket{g_n}$ is 
properly chosen, the resulted Wannier function is maximally localized. No further 
minimization such as the one in Ref.~\cite{Marzari12} is needed.

\begin{figure}[ht]
\includegraphics[width=0.8\linewidth]{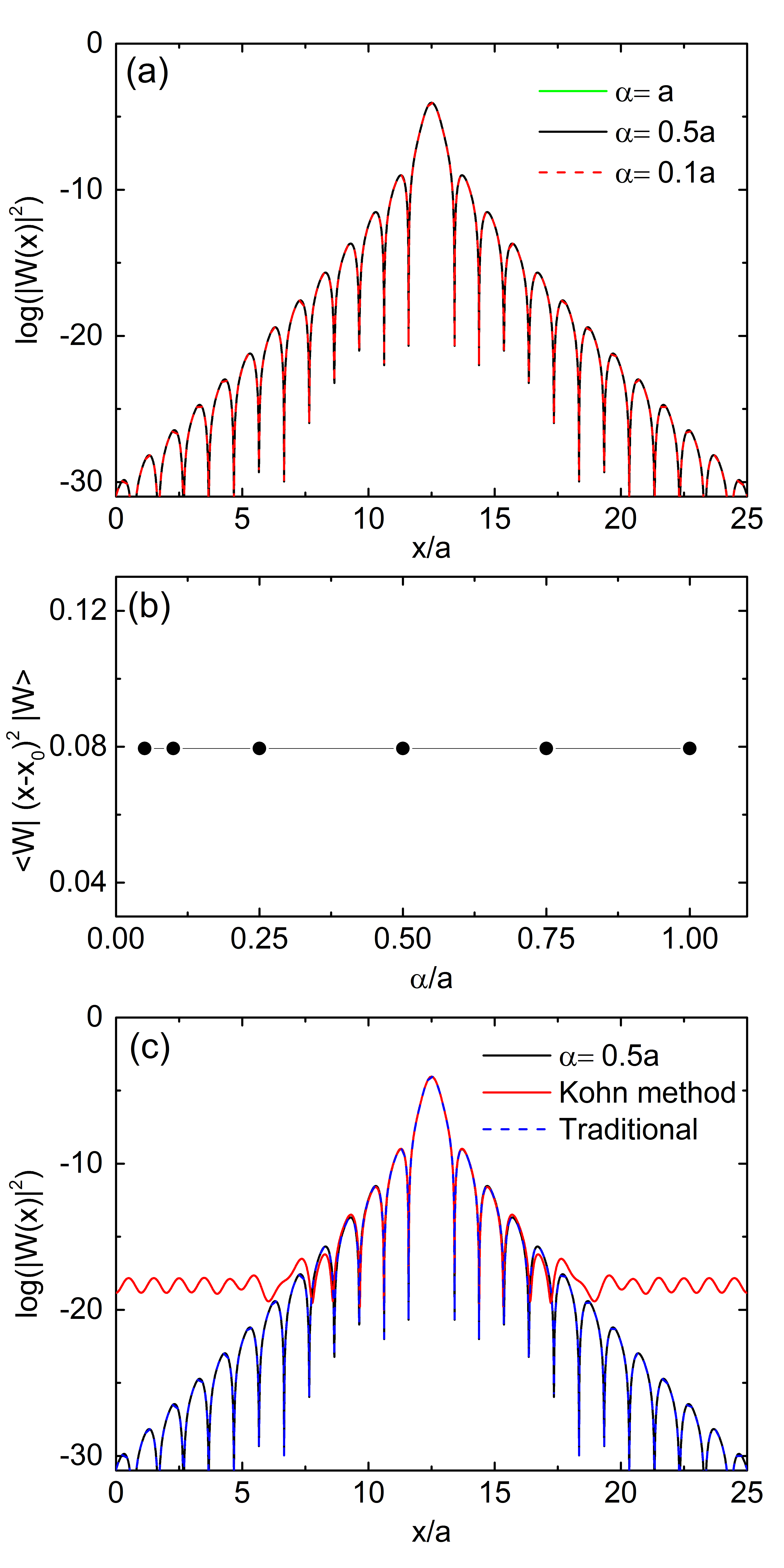}
\caption{(color online) Wannier functions for one dimensional periodic potential $V(x)=\cos(2\pi x/a)$.  (a) The Wannier functions obtained with our method with trial Gaussian functions
of different widths $\alpha$. (b) The spread of the Wannier function obtained with our method as 
a function of the width of the Gaussian function $\alpha$. (c) The Wannier functions obtained
with three different methods. The black line is our method; the blue line 
is obtained with the traditional method, where one fixes the phases of Bloch wave
according to the prescription given in Ref. ~\cite{Kohn59}; the red one is
obtained by orthonormalizing $\ket{\xi_n}$'s with Kohn's method~\cite{Kohn1973orth}.
}
\label{ssband}
\end{figure}

To illustrate our method, we consider a single particle 
Hamiltonian with the periodic potential $V(x)=\cos(2\pi x/a)$, where $a$ is the 
lattice constant. We construct the Wannier function for its lowest band.
As each potential well is symmetric with respect to its center (the
lowest point of the well),  we expect that the MLWF is also symmetric with
its highest peak at the center. So, we  choose 
$\braket{x|g_n}=e^{-(x-na)^2/2\alpha^2}/\sqrt{2\alpha^2\pi}$. 
It is clear that $\braket{w_n | g_n}$ is the largest
for a narrowest Wannier function allowed within the band or MLWF. The resulted
Wannier function should be insensitive to the width of $\braket{x|g_n}$.
This is exactly what we see in Fig. \ref{ssband}(a). 
With three Gaussians of different widths $\alpha=a, 0.5a, 0.1a$ as trial functions, 
the resulted Wannier functions fall right on top of each other 
as seen in Fig. \ref{ssband}(a).  We have computed the spread of the Wannier functions
obtained with our method; they are almost identical for  Gaussian trial functions of 
different widths as seen Fig. \ref{ssband}(b).  

To further illustrate the effectiveness of our method,
we have computed Wannier functions with two other methods. One
is the traditional method, where one fixes the phases of the Bloch waves
according to the prescription in Ref. \cite{Kohn59}. The other method 
is to orthonormalize $\ket{\xi_n}$'s with Kohn's method~\cite{Kohn1973orth}. 
They are compared to our results in  Fig. \ref{ssband}(c), where we 
see that our result is in excellent agreement with the traditional method while
the one obtained with Kohn's method is much worse.


The  L\"owdin method can be implemented differently. However,  as
long as it is implemented correctly,  the method  transforms a given set
of non-orthogonal vectors to a unique set of orthonormal vectors.  Nevertheless,
we show here explicitly how we implement it. We impose a periodic boundary
condition with $N$ unit cells. As a result, the crystal wave vector $k$ takes $N$
discrete values ${k_1, k_2, \cdots, k_N}$.
We let $\ket{\psi_j}=\ket{\psi_{k_j}}$ and $A_{nj}=\braket{\psi_j|g_n}$.
The L\"owdin orthonormalization is then implemented as
\be
\ket{w_n}=\sum_{mj} (AA^\dagger)^{-1/2}_{nm}A_{mj}\ket{\psi_j}\,.
\label{lowdin}
\ee
If the trial function $\ket{g_n}$ is translationally symmetric, $\braket{x|g_n}=\braket{x-r_n|g_0}$,
we have $A_{nj}=e^{-ik_jr_n}A_{0j}$ and
$(AA^\dagger)_{nm}=\sum_j e^{ik_j(r_m-r_n)}|\braket{\psi_j|g_0}|^2$.

\begin{figure}[ht]
\includegraphics[width=0.85\linewidth]{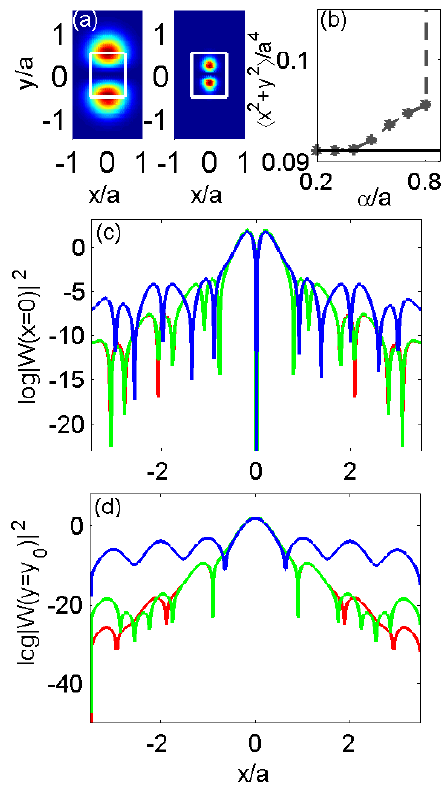}
\caption{\label{fig:wanner2D} (color online) Wannier functions of the composite $p$-bands
in a two dimensional lattice $V(x,y)=V_0[\cos(2\pi x/a)+\cos(2\pi y/a)]$ with $V_0=-30$.
(a) The left is the trial wave function $g_2$ with $\alpha=0.5a$;
the right is the Wannier function obtained with this trial function. The white dashed line 
marks the unit cell.  (b) The spread
$\langle w| x^2+y^2 | w\rangle$ of a Wannier function with respect to varying $\alpha$. 
The convergence to the black line is obvious.  The black line 
is the width of the Wannier function $w(x,y)=w_1(x)w_2(y)$, where 
$w_1(x)$ as the $s$-band Wannier function of $\cos(2\pi x/a)$ and $w_2(y)$ as the $p$-band
Wannier function of $\cos(2\pi y/a)$ are obtained with the traditional method. 
(c) Logarithmic plot of Wannier functions at $x=0$ along $y$-axis for different values of $\alpha$: $\alpha=0.3a$(red), $0.6a$ (green), $0.88a$ (blue). Note that the latter two almost overlap.
(d) Logarithmic plot of Wannier functions at $y_0$ along the $x$-axis for different choices of $\alpha$: $\alpha=0.3a$(red), $0.6a$ (green), $0.88a$ (blue).
$y_0$ is the highest peak position of the Wannier functions.  }
\end{figure}

Our method is applicable for 
composite bands. In composite bands, one or more Wannier functions have
nodes. Therefore, to have largest $\braket{w_n | g_n}$ for MLWFs, 
we need to choose such $\ket{g_n}$'s that they have nodes at proper positions. 
The node positions can be determined by symmetries of the wells. In the worst case, 
we can determine these node positions by numerically computing the eigenstates
of the local wells. Here we consider a two dimensional periodic potential
$V(x,y)=V_0[\cos(2\pi x/a)+\cos(2\pi y/a)]$ and use its p-bands to illustrate the effectiveness of
our method for composite bands. The two trial wave functions are chosen as 
$g_{1,2}=\sqrt{\frac{\pi}{2\alpha^4}} \eta_{1,2} e^{-(x^2+y^2)/2\alpha^2}$ 
with $\eta_1=x, \eta_2=y$, which are the two first excited states  of a two dimensional 
harmonic oscillator.  The results are plotted 
in Fig.\ref{fig:wanner2D}. Shown in Fig.\ref{fig:wanner2D}(a) is a comparison 
between the trial function
and the resulted Wannier function. Three Wannier functions obtained from different trial functions
are shown in Fig.\ref{fig:wanner2D}(c) and (d). Our numerical calculations show 
that when the width of the trial function
$\alpha$ is narrow enough, the resulted Wannier functions are almost identical to each other. 
When they are plotted in the figure, they fall right on top of each other. So, in 
Fig.\ref{fig:wanner2D}(c) and (d) only three Wannier functions are plotted. 
The spread of a Wannier function $\langle w | x^2+y^2 | w \rangle$ is plotted as 
the function of $\alpha$ in Fig.\ref{fig:wanner2D}(b),  which shows that the Wannier function
spread no longer changes when $\alpha$ is narrow enough. 

Our method is directly applicable to random potentials. 
The reason is that we use the full band projection. It can  be constructed
with all the energy eigenfunctions in a given band no matter 
the eigenfunctions are Bloch waves or not. It is well known that 
even in random potentials, there exist eigen-energy ``band" that are isolated 
from other eigen-energies by gaps that are independent of the system size.  
For these bands, there exist Wannier functions~\cite{Nenciu93}. 
Our method can be used to compute
these Wannier functions by constructing the projection $P$
with the energy eigenstates of a given random energy band. According to Eq.(\ref{sum}), 
the resulted Wannier functions maximize a sum, $\sum_n \braket{w_n|g_n}$.  
With a proper choice of
$\ket{g_n}$'s, these Wannier functions can be regarded as maximally localized collectively.
There already exist several methods to compute Wannier functions for random potentials. 
Our method has various advantages. The method in Ref.~\cite{Kohn73} is only 
applicable to the case where the random potential is a perturbation to a  
periodic potential. Kivelson's method~\cite{Kivelson82} has difficulty for two or three dimensional systems. Our method
is clearly more efficient than the one in Ref.~\cite{Ceperley10}. We ourselves have recently
proposed a method to compute Wannier functions for random potentials~\cite{Zhu2016}.
Our current method is certainly superior.

\begin{figure}[ht]
\includegraphics[width=0.9\linewidth]{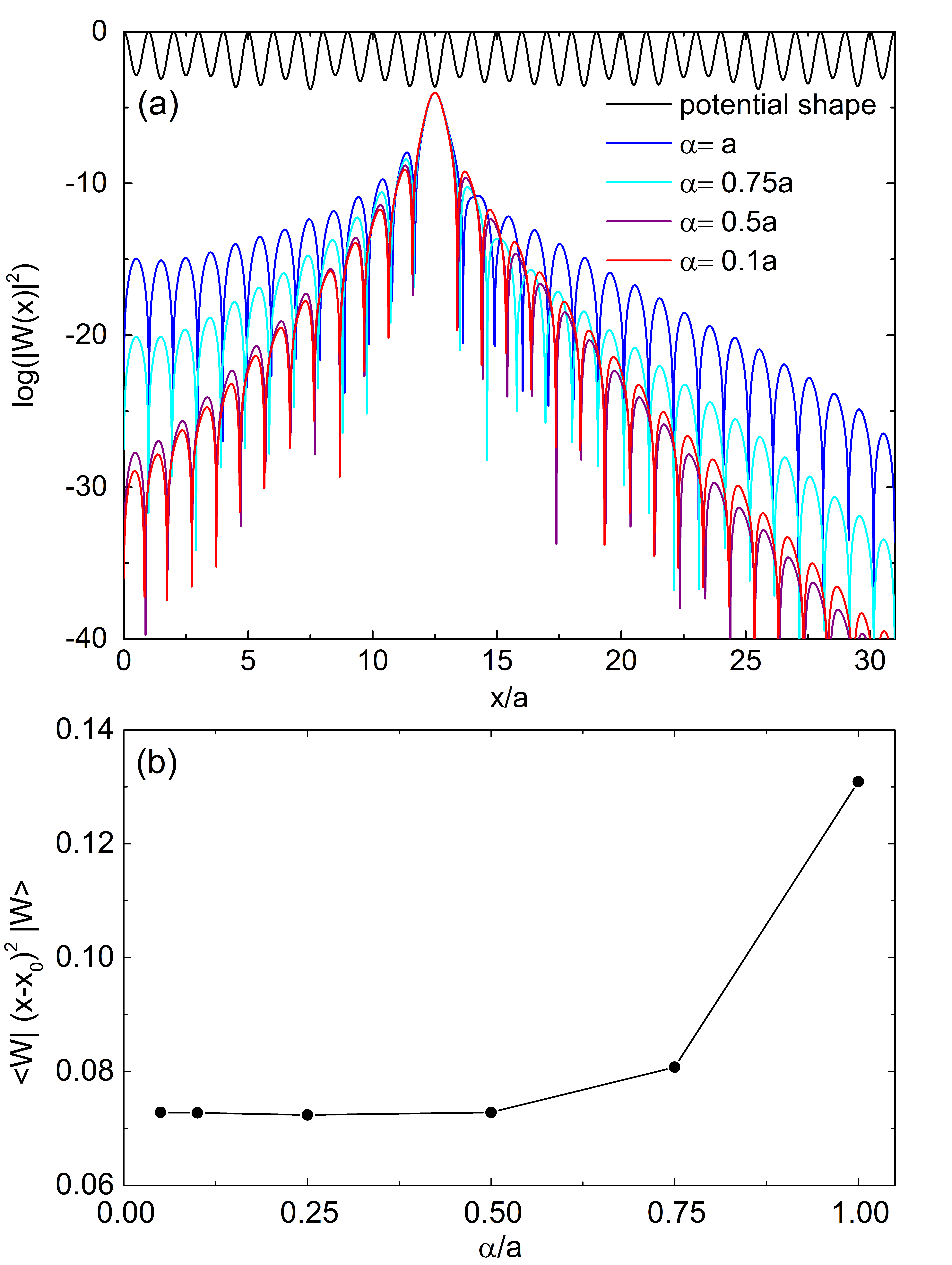}
\caption{(color online) Wannier functions constructed by our method for one dimensional 30\% disordered potential, which is plotted as the black line at the top of (a).  
The initial trial Gaussian functions
are the same for different wells. (a) The Wannier functions obtained
with Gaussian functions of different widths $\alpha$. (b) The spread of the 
Wannier functions obtained with our method as a function of the width of
the trial Gaussian function. }
\label{disordered}
\end{figure}

We now  illustrate  our method in disordered systems.
We choose a disordered potential as a series of cosine-type wells
of random depths, $V_n(x)=A_n[\cos(2\pi x/a)-1]$. Well depths $A_n=A[1+\eta\cdot\mathfrak{R}_n]$, 
where $A$ is a constant and $\mathfrak{R}_n$ denotes a sequence of random numbers
between -0.5 and 0.5, and $\eta$ denotes the relative strength of disorder. 
For instance, we refer to $\eta=0.1$ as a 10\% disorder. We use $A=5$ and $a=1$, and $\eta=0.3$ 
in the example shown in Fig. \ref{disordered}. In our computation, we choose
the Gaussian trial functions $\ket{g_n}$ for different wells despite the wells are different. 
As shown in Fig. \ref{disordered},  our method produces successfully  exponentially
localized Wannier functions. It is clear that  a narrower 
trial Gaussian function leads to more localized Wannier functions. Numerical results indicate that Gaussian trial functions 
with $\sigma$ as small as 0.5$a$ is enough to construct MLWFs.

For bands of non-trivial topology, the existing methods~\cite{Marzari97,Teichler}
face inherent singularity at special $k$ points. One has to find ingenious ways to 
circumvent the singularity to obtain exponentially localized Wannier 
functions~\cite{Winkler2016}. Our method has the potential to circumvent 
the singularity without any special re-design
as we use the full band projection instead of the projection at individual $k$ points. 

Note that we have so far assumed that the system has time reversal symmetry. For these systems,
its energy eigenfunctions can always be made real. It follows that the projection $P$ and all the matrices involved in  the L\"owdin method (Eq.\ref{lowdin}) can also be made real.
Therefore, the final Wannier functions $\ket{w_n}$ are also real as long as 
the trial functions are real. For systems where the time reversal symmetry 
is broken, $\ket{w_n}$ can be complex. In this case, the L\"owdin method 
maximizes ${\rm Re}\{\braket{w_n|g_n}\}$ and we may not obtain
MLWFs. We shall leave it for future discussion.

In sum, we have presented a simple and general method for constructing MLWFs.
In our method, the full band projection is used on localized trial wave functions.
If the trial functions are properly chosen to respect the local potential configuration and have good
node positions, the ensuing L\"owdin method orthonormalize them to MLWFs.
No numerical minimization is needed. Our method can be directly applied to random potentials.
The application of our method to a real material is left for future.

We thank Ji Feng and Xianqing Lin for helpful discussion. This work is supported by the National Basic Research Program of China (Grants No. 2013CB921903 and No. 2012CB921300) and the National Natural Science Foundation of China (Grants No. 11274024, No. 11334001, and No. 11429402).


\end{document}